# 150 Years of Return Predictability Around the World:
# A Holistic View Across Assets[†]

Yang Bai

University of Missouri

(Preliminary Draft)

First Draft: Aug. 19, 2022

This Draft: Aug. 19, 2022

## Abstract

Using new annual data of 16 developed countries across bond, equity, and housing markets, I study the return predictability using the payout-price ratios, i.e., coupon price, dividend price, and rent price. None of the 48 country-asset combinations shows consistent in-sample and out-of-sample performance with positive utility gain for the mean-variance investor. Only 3 (4/2) countries show positive economic gains in their equity (housing/bond) markets. The return predictability for the representative agents' risky asset portfolios and wealth portfolios is even weaker, suggesting that timing the investment return of a country using payout-price ratios will not make the investors better off. The predictive regressions based on the VAR analysis by Cochrane (2008, 2011) suggest that 14 (5) countries have predictable payout growth in the equity (housing) markets, ex., the dividend price predicts the dividend growth in the US. The VAR simulation using data from all the countries does not reject the null that the dividend growth is predictable. This paper presents firm evidence against the return predictability based on payout ratios.

**Keywords:** Bond Return Predictability, Certainty Equivalent Return, Coupon Price, Dividend Price, Dividend Predictability, Equity Return Predictability, Housing Return Predictability, Out-of-sample $R^2$, Rent Price

**JEL Classification:** G10, G11, G12, G14

---

# 1.    Introduction

Campbell and Shiller (1998b, a) derive a definition of returns in the following form:

$$d_t - p_t \approx const. + E[\Sigma_{j=1}^{\infty} \rho^{j-1}(r_{t+j} - \Delta d_{t+j})],$$

where $d_t - p_t$ is the dividend-price ratio, $r_{t+j}$ is the return, and $\Delta d_{t+j}$ is the dividend growth. Because dividend is persistent, the definition highlights the time-varying return driven by payout-price ratios, leading to the conclusion that the return must be predictable. Therefore, the majority of the return predictability literature devotes its focus to dividend-price ratio and the equity market returns, specifically with the US data post WWI (Goyal and Welch, 2003; Cochrane, 2008). However, the US equity market in the past 100 years reveals only one of the infinitely many trajectories that the return series can take, and the above relation applies not only to the equity markets but also to all the other markets with time-varying payout-price ratios, including bond and housing markets.

The recent literature shows that the conclusions from the US equity market since 1927 or the 1950s may not generalize to other countries and time. For example, contrary to the earlier findings in the literature on dividend predictability, Rangvid, Schmeling, and Schrimpf (2014) and Chen (2009) show that the conclusions of the dividend predictability based on the US equity market do not directly generalize to the equity markets of other countries nor subperiods. Golez and Koudijs (2018) also show that dividend price predicts returns by extending data backward in time. Studies of return predictability of bonds and housing markets often do not take on the perspective of payout-price ratios, leading to limited interpretation for the universal relation based on Campbell-Shiller decomposition (Yang, Long, Peng, and Cai 2020; Gargano, Pettenuzzo, and Timmermann, 2017; Cochrane and Piazzesi, 2005; Campbell and Shiller, 1991, 1988b, a). Collectively, the literature calls for an examination for the return predictability across assets and countries primarily to answer whether the return predictability exists based on payout price ratios. This paper takes this challenge and examines return predictability on a large scale.

With a new database including 16 countries and three different asset markets, I provide a holistic view on the return predictability of payout-price ratios across asset classes and countries from the 1870s to 2020. I include three asset classes: long-term government bond, equity, and housing. I also consider the aggregated portfolios to shed a light on the total returns to the representative agent in each country.

Despite recent debate over return predictability, I confirm that the hope of return predictability based on payout-price ratios is slim across assets globally (Goyal and Welch 2021; Golez and Koudijs 2018). Table 1 reports the summary of the findings. Consistency in return predictability evaluated in three aspects, i.e., in-sample (IS) significance, out-of-sample (OOS), and OOS utility gains, is a universal problem for all the asset classes, including bond, equity, and housing assets. The out-of-sample performance is terrible. However, predictive regressions based on Cochrane's VAR analysis show that 14 countries have predictable dividend growth and that 5 countries have predictable rent growth[1] (Cochrane 2008, 2011).

**[INSERT TABLE 1 ABOUT HERE]**

Of the 48 country-asset combinations, no country-asset combination shows consistent predictability. The bond excess returns are the most difficult to predict with no meaningful OOS predictability, while the housing excess returns are marginally easier to predict out of the sample. The housing market shows positive CER gains 4 out of 16 cases. However, 2 out of these CER gains are very close to zero. Given the transaction costs, ex., commission alone can be as high as 6% in the US, whether timing the housing market with rent growth can be economically meaningful — even for the most promising market, i.e., the Finnish housing market with 0.61% CER gain — remains a question.

With risky asset portfolios based on the value-weighted average of equity and housing returns and the wealth portfolio based on the value-weighted average of equity, housing, bond, and treasury bill returns, I also examine the prediction of the countries' representative agents' portfolios, from which we can learn about the likelihood that an investor can benefit from timing the overall asset returns of a country. These tests can contribute to our understanding of the predictability over the total returns, without considering to which market the theoretical relation should apply. The results are still disappointing. Both the risky asset portfolios and the wealth portfolios are not predictable if we account for the OOS performance. It is hard to realize anything through timing a country's overall asset returns.

---

[1] The bond payout growth is 0, i.e., $\Delta d = 0$, because coupon rate is fixed. Therefore, the predictive regressions do not include the bond markets.

Cochrane (2008, 2011), among others, points out that if return predictability does not exist, dividend growth must be predictable. With the new data covering the longest time window globally in the literature, the ordinary least square regressions show that dividend growth is predictable in 14 countries and that rent growth is predictable in 5 countries, emphasizing the absence of predictability from a theoretical perspective. Consistent with the findings from the US equity market post WWII (Golez and Koudijs, 2018; Chen, 2009), from 1873 to 2020, dividend growth is also predictable even in the US equity market, although the returns of the US equity market are not predictable during the same period. The simulation following Cochrane (2008) also do not reject the joint hypothesis that the returns are not predictable and the dividend growth is predictable. These findings present firm evidence against what the previous literature documents (Cochrane, 2008).

This study contributes to the literature in three aspects. First, it expands the understanding of return predictability across the asset markets and provides the *first* holistic view under the theoretical payout-price perspective of return predictability in the bond, the equity, and the housing markets. In the previous literature, bond return predictability and the housing return predictability are absent in the return predictability studies using payout-price ratios. However, these assets take up important percentage in the representative agent's portfolios. Housing markets, particularly, are of central importance to the investors globally. As Jordà, Knoll, Kuvshinov, Schularick, and Taylor (2019) point out, "housing has been a good long-run investment as equities and possibly better". More importantly, Campbell-Shiller decomposition does not require an assumption of asset classes (Campbell and Shiller, 1988b, a).

Second, the studies on the representative agents' portfolios were not possible before this new data. However, the representative agents' asset holdings are important to the literature. With this new data, we have a good definition of the aggregated risky asset market and a good definition of the aggregated wealth market including everything. Through the empirical findings using the representative agents' portfolios, this paper contributes to the understanding of the relation between the payout price and the returns in the theory that often does not specify to which market the relation between payout price and return should apply. It points out that the return predictability is difficult to realize for the aggregated asset portfolios at the country level.

In the end, it introduces to the finance literature a new macrohistory data with broad coverage and updates the literature's understanding of the return predictability by focusing on dividend variables with long global historical data extending to the 1870s. For example, it rules out the likelihood that the dividend growth in the US equity market is not predictable over the long run.

Across assets and countries, the findings from this paper provide firm evidence that questions the return predictability implied by Campbell-Shiller (1988b, a) decomposition based on payout-price ratios, offering unique insights to the literature.

The paper now proceeds as follows. Section 2 provides a brief description of the database that I adopt and the empirical methods for in-sample and out-of-sample tests. Section 3 provides the main results, including the discussions on in-sample tests with linear regressions, out-of-sample $R^2$, and economic performance measured in the mean-variance investors' utility gains. Section 4 concludes the paper.

## 2.      Data and Empirical Methods

I first provide a brief description of the database I adopt and the markets I analyze. I adopt the new Jordà-Schularick-Taylor macrohistory database developed by Jordà, Knoll, Kuvshinov, Schularick, and Taylor (2019)[2].

I focus on the nominal returns and the payout-price ratios, including coupon-price, dividend-price, and rent-price ratios. The tests are conducted against long-term government bonds, stocks, and housing assets at the country level. Because the database only provides coupon yield for the bond markets, i.e., $\frac{c_t}{p_{t-1}}$, I back out the coupon price defined as $\frac{c_t}{p_t}$ with the provided bond returns. The other payout-price ratios, i.e., dividend-price and rent-price ratios, are provided by the database.

For each country, I calculate the excess returns of each market and the excess returns of the representative agent's portfolios. Then, I create the lagged payout-price ratios. Following the literature, I define both the payout-price ratios and the excess returns in log scales. The housing prices and the rents are generally defined as the average housing prices and the average rents for the country. The bond returns, as well as the asset allocation practice, assume annual rebalance on new long-term government bonds[3]. Table A1 reports the summary statistics.

In general, the returns are representative (Jordà, Knoll, Kuvshinov, Schularick, and Taylor 2019). For example, the authors of the database validate the aggregation of the countries level housing market returns using different methods and from different perspectives.

## 2.1    Empirical Methods

This paper focuses on the three popular metrics in the literature, i.e., IS significance measured mainly with the t statistics, OOS $R^2$, and CER gains. For the IS tests, I adopt the predictive ordinary least squares regression with lagged payout-price ratio.

---

[2] The Jordà-Schularick-Taylor macrohistory database is maintained by the University of Bonn, Germany. See https://www.macrohistory.net/database/ for details. This paper adopts the 6th update of the database.
[3] The maturities vary. See Jordà, Knoll, Kuvshinov, Schularick, and Taylor (2019) for details.

$$r_{i,j,t}^e = a_{ij} + b_{i,j}\, dp_{i,j,t-1} + \varepsilon_{i,j,t},$$

where $dp_{i,j,t-1}$ stands for the payout-price ratio for asset I in country j in year t-1. In other words, for each asset in each country, I separately fit the country-asset combination an IS regression. With the aggregated portfolios, i.e., the representative agent' portfolios in each country, I include all three payout-price ratios, i.e., coupon price, dividend price, and rent price. I report Newey-West t statistics for the coefficients.

In the OOS tests, I start modeling with at least 20 years and adopt an expanding window that uses all the past observations. I roll forward this setup to make one-year-ahead predictions in each year. The predictions are then evaluated based on the OOS $R^2$ statistics. Following the literature, I define OOS $R^2$ as:

$$R_{i,j}^2 = 1 - \frac{\Sigma_{t=T_1}^T \left(r_{i,j,t}^e - \hat{r}_{i,j,t|t-1}\right)^2}{\Sigma_{t=T_1}^T \left(r_{i,j,t}^e - \bar{r}_{i,j,t|t-1}\right)^2},$$

where $T_1$ is the starting year of prediction, $\hat{r}_{t|t-1}$ is the alternative prediction delivered by payout-price ratio, $\bar{r}_{i,j,t|t-1}$ is the prevailing historical average of excess return, and $r_{i,j,t}^e$ is the realized true excess return (See Goyal and Welch 2008; Campbell and Thompson 2008; Nealy, Rapach, Tu and Zhou 2014).

Besides statistical metrics, I also report OOS economic performance measured in CER gain calculated as the difference between the CER delivered by the portfolio based on the alternative predictions and the CER delivered by the portfolio based on the prevailing historical average of excess returns. Following Brennan and Xia (2004), Campbell and Thompson (2008) and Goyal and Welch (2008), I define CER as:

$$CER = E[R_p] - \frac{\gamma}{2} \cdot Var(R_p),$$

where $R_p$ is the portfolio of mean-variance investor with risk aversion coefficient $\gamma = 5$. The portfolio is formed for the mean-variance investor through Markowitz rule. The mean-variance investor allocates her fund between the asset of interest, i.e., long-term government bond, equity, and housing, and the risk-free rate proxied by treasury bill of the local market. The weight on risky assets is:

$$w_{i,j,t} = \frac{1}{\gamma} \frac{\hat{r}_{i,j,t|t-1}}{\hat{\sigma}_{i,j,t|t-1}},$$

where $\hat{r}_{i,j,t|t-1}$ is the predicted excess return in percentage and $\hat{\sigma}_{i,j,t|t-1}$ is the predicted variance of excess returns (Campbell and Thompson 2008). I estimate the variance with the past 20-year realized variance. The mean-variance investor decides the weight $w_t$ to allocate the fund to the risky assets. Following the literature, I force the weight to be in [0,1.5], i.e., the investor can take a 50% leverage in her position but cannot short sell the asset (Campbell and Thompson 2008; Goyal and Welch 2008; Nealy, Rapach, Tu, and Zhou 2014).

To understand the trading cost difference between portfolios based on the prevailing average of excess returns and the portfolios based on the predictions, I calculate the relative turnover ratio using the weight changes in the mean-variance investor's portfolio. The relative turnover ratio is calculated as the turnover rate of the portfolio based on the prevailing average of excess returns over the turnover rate of the portfolio based on the predictions. The turnover rate is defined as:

$$Turnover = \frac{1}{T} \Sigma_{t=1}^{T} \Sigma_{k=1}^{2} (|w_{i,j,k,t+1} - w_{i,j,k,t}|),$$

where $T$ is the number of years in OOS predictions and $w_{i,j,k,t+1}$ is the weight for asset I in country j. The subscript k stands for treasury bill when k=1 and stands for risky assets, i.e., one of bond, equity, and housing, when k=2 (DeMiguel, Garlappi, and Uppal 2009; Nealy, Rapach, Tu, and Zhou 2014).

Towards the end of this paper, I conducted another set of predictive regressions for the payout growth in the equity and the housing markets using payout-price ratios, i.e., dividend price and rent price. Specifically, I fit the following ordinary least squares regression:

$$\Delta d_{i,j,t} = a_{i,j} + \beta_{i,j} dp_{i,j,t-1} + \varepsilon_{i,j,t},$$

where $\Delta d_{i,j,t}$ is the payout growth. This regression is implied by Cochrane's VAR analysis (Cochrane 2008, 2011).

Cochrane points out that if the return predictability does not exist, the dividend growth should be predictable, i.e., the test on the return predictability is a joint test of the return predictability and the dividend predictability. He illustrates that the dividend price, dividend growth, and return can have systematic relations as the following VAR system:

$$\begin{bmatrix} d_{t+1} - p_{t+1} \\ \Delta d_{t+1} \\ r_{t+1} \end{bmatrix} = \begin{bmatrix} \phi \\ b_d \\ b_r \end{bmatrix} (d_t - p_t) + \begin{bmatrix} \varepsilon_{t+1}^{dp} \\ \varepsilon_{t+1}^{dp} \\ \varepsilon_{t+1}^{r} \end{bmatrix}.$$

Through economic identities based on Campbell-Shiller decomposition (1988b, a), this system reduces to a bivariate VAR system including the bottom two relations. When we test the return predictability through the coefficient $b_r$, we cannot ignore the intrinsic relation between $b_r$ and $b_d$. He argues that, instead of testing return predictability, testing dividend growth can provide a more convincing conclusion. I follow Cochrane (2009) and include his simulation tests towards the end of section 3. Because the coupons on long-term government bonds are usually fixed, i.e., $\Delta d = 0$, I focus on the equity and the housing markets for both the predictive regressions and the VAR simulation.

## 3.    Empirical Results

I report the results from the IS predictive Regressions in this section. Using lagged payout-price ratios, I fit OLS regressions using excess returns. In summary, the IS regressions show limited predictability across countries for the bond and the equity markets. Payout prices predict excess returns in only two countries' bond markets and five countries' equity markets. However, the housing market results are compelling. The rent-price ratios predict excess returns in 8 countries' housing markets, demonstrating a promising potential for consistent predictability. Table 2 reports the results.

**[INSERT TABLE 2 ABOUT HERE]**

Besides the predictions made for the asset markets directly, I also seek to understand whether we can make successful predictions for each country's representative agent's portfolios. Using the unique database, I calculate excess returns for the risky asset portfolios and the wealth portfolios in each country. Each risky asset portfolio comprises the equity and the housing markets, weighted by their respective capitalization. Similarly, each wealth portfolio comprises the bond market, equity market, the housing market, and the treasury bill market, weighted by their respective capitalization. Including all three payout-price ratios, the representative agents' portfolios seem promising in presenting strong return predictability according to the IS regressions. Only three countries have no IS predictability for risky asset portfolios, i.e., none of the three payout-price ratios has a significant regression coefficient. 12 wealth portfolios show the potential of realizing return predictability based on the IS tests.

The OOS tests are much tougher. Table 3 reports the results. Based on the OOS $R^2$ statistics, no OOS predictability exists in any of the 16 bond markets. This finding is consistent with the findings from the US (Thornton and Valente, 2012). Dividend price predicts only one equity market out of the sample: the UK equity market. Despite strong evidence in the IS tests, the housing markets perform no better. Only two housing markets seem predictable: the French housing market and the German housing market.

**[INSERT TABLE 3 ABOUT HERE]**

Consistency between the IS performance and the OOS performance does not exist. No bond market out of the 16 countries shows performance that is consistent in the sample and out of the sample. Only the British market show hope for equity. Housing markets are similar, and the two markets that show some potentials are the French housing market and the German housing market. For the representative agents' portfolios, only the French representative agent enjoys consistent IS and OOS predictability for the risky asset portfolio and the wealth portfolio.

I evaluate the economic performance for the predictions in the OOS periods through CER gains, the utility gains of the mean-variance investors in each country. Table 4 reports the findings of the economic performance. The economic performance in the bond and the equity markets is consistent with the weak IS and OOS performance. Two countries show positive CER gain in the bond markets, while the equity markets in Australia, Netherland and Spain show positive CER gains. Despite the strong IS performance in the housing markets, the OOS economic performance in the housing markets is consistent with the disappointing OOS statistical performance. Only four housing markets deliver positive CER gains, and two of the four markets show very limited CER gains. Note that the transaction costs in the real estate market are huge, and the mean-variance investors from all the countries will probably forgo even the positive CER gains when they consider the transaction costs[4].

[INSERT TABLE 4 ABOUT HERE]

Finally, table 4 also reports the trading turnover measured as the relative ratios (See section 2). Although the economic performance is weak, the turnover rates based on the payout-price ratios in the bond and the equity markets are higher than the turnover rates of the portfolios based on the prevailing excess returns by multiple folds. For example, in the Italian equity market, despite realizing nonpositive CER gains and sizably lower Sharpe ratios, the turnover rates of portfolio based on the dividend-price ratios can be 4 times higher than the turnover rates of the portfolios based on the prevailing excess returns. The turnover rates of the housing portfolios can also be several times higher than their corresponding null portfolios' turnover rates. In the French housing market, trading for the prediction portfolio is about 4 times as frequent as the trading for the portfolio based on the prevailing excess returns. Consistent with the CER results, it is unlikely that the strategies based on rent-price ratios will generate any profits out of the sample, even if we do not consider the high transaction costs in the housing markets. The representative agents' portfolios show much worse turnover rates. Many countries show negative economic gains with relative turnover ratios above 5, emphasizing the fact that the mean-variance investors in these countries cannot get better off economically by timing the representative agents' portfolio returns.

---

[4] For example, see https://homebay.com/tips/real-estate-agent-commission-101-for-sellers/. The commission fee alone can be close to 6% in the US. A careful study from Jordà, Knoll, Kuvshinov, Schularick, and Taylor (2019) further indicates that the transaction costs in 13 out of the 16 countries in the housing markets are around 7.7% of the property's value. Despite the frequent trading can incur high cumulative costs in the bond and equity markets throughout a year, this paper considers annual rebalance, and thus the transaction costs in the housing markets are much higher than the transaction costs in the equity markets and bond markets. Although the taxes are considered in all the markets, Jordà, Knoll, Kuvshinov, Schularick, and Taylor (2019) documents that it is unlikely that the taxes can contribute to the differences of the returns in these markets.

Because of the unique relation between dividend growth and return predictability, Cochrane (2008, 2011) argues that if the return predictability does not exist, the dividend predictability should be observable, emphasizing the need to test the predictive relation between payout-growth and payout-price ratios. I test the relation with ordinary least squares regressions across the countries and the markets. Table 5 reports the results.

**[INSERT TABLE 5 ABOUT HERE]**

Consistent with the findings in the 50-year *quarterly* data *from 1973 to 2009* by Rangvid, Schmeling, and Schrimpf (2014), 14 countries show significant loadings on dividend price in their equity markets in the past 150 years, including the US (See also Rangvid, Schmeling, and Schrimpf, 2014). This finding is contradictory to the previous findings in the literature for long term divided predictability (ex., Chen 2009). The relation between the payout-price and the excess return also exists in five housing markets, including Belgium, Germany, Portugal, Switzerland, and the US. If the theory is correct and we rely on the IS tests of the return predictability and the payout predictability, we can safely conclude that the return predictability does not exist in the equity markets of Denmark, Finland, Germany, Italy, Japan, Netherland, Norway, Spain, Sweden, Switzerland, and the US, and the return predictability does not exist in the housing markets of Belgium, Switzerland, and the US.

To supplement the IS tests for the payout predictability, I also conduct the simulation of the VAR system including the dynamic relations of the payout price series, the payout growth series, and the return series (Cochrane 2008). Cochrane (2008) argues that the tests on return predictability is a joint test involving three parameters, i.e., $\phi$, $b_d$, and $b_r$, in the following VAR system:

$$\begin{bmatrix} d_{t+1} - p_{t+1} \\ \Delta d_{t+1} \\ r_{t+1} \end{bmatrix} = \begin{bmatrix} \phi \\ b_d \\ b_r \end{bmatrix} (d_t - p_t) + \begin{bmatrix} \varepsilon_{t+1}^{dp} \\ \varepsilon_{t+1}^{dp} \\ \varepsilon_{t+1}^{r} \end{bmatrix}.$$

However, based on Campbell-Shiller (1988b, a) decomposition, two identity relations exist that reduce one dimension of the system, leading to the VAR system below under the null hypothesis that return predictability does not exist, where the parameters reflect the joint null hypothesis, i.e., $b_r = 0$, and $b_d = \rho\phi - 1 < 0$.

$$\begin{bmatrix} d_{t+1} - p_{t+1} \\ \Delta d_{t+1} \\ r_{t+1} \end{bmatrix} = \begin{bmatrix} \phi \\ \rho\phi - 1 \\ 0 \end{bmatrix} (d_t - p_t) + \begin{bmatrix} \varepsilon_{t+1}^{dp} \\ \varepsilon_{t+1}^{d} \\ \varepsilon_{t+1}^{d} - \rho\varepsilon_{t+1}^{dp} \end{bmatrix}.$$

Under the assumptions, Cochrane (2008) argues that if the return predictability does not exist, then dividend growth has to be predictable, and $b_d$ has to be equal to $\rho\phi - 1 < 0$. Using data from all the markets, I conduct the VAR simulation analysis with 10000 repetitions following Cochrane (2008) for the equity and the housing markets. Figure 1 reports the findings.

**[INSERT FIGURE 1 ABOUT HERE]**

The simulation does not reject the null hypothesis that the returns are not predictable in the equity and the housing markets. However, different from the findings in other studies, the hypothesis that dividend growth is predictable is also not rejected in the equity markets using all of the international data, presenting new evidence against the time series return predictability globally (Cochrane 2008; Golez and Koudijs 2018). On the other hand, the simulation suggests that there is some hope to observe return predictability in the housing markets. However, since the observed coefficient is negative, it is possible that the true coefficient for rent price is indeed negative, leaving a low chance to observe a positive $b_r$.

## 4. Conclusion

Return predictability has been a central focus of the asset pricing literature, because it is regarded as a crucial indicator of time-varying returns. Although the literature reaches some solid conclusions for the OOS performance with the US equity market data (Goyal and Welch, 2008, 2020), a paucity of research into the generalizability of such conclusions exists for other asset markets and across countries, where the theoretical conclusion also applies, and the predictability in the *long run* for payout growth, ex., dividend growth, remains inconclusive globally.

Specifically, despite the similarity, no study has documented the return predictability of payout-price ratios in the bond and the housing markets with a holistic view. In the recent years, a couple papers also point out the possibility that the findings of the US equity market cannot generalize to the other (sub) time periods and countries (Golez and Koudijs, 2018; Rangvid, Schmeling, and Schrimpf, 2014; Chen, 2009). This paper addresses these issues through a comprehensive examination across assets and countries.

I provide to the literature the *first* holistic view on the return predictability implied by Campbell and Shiller (1988b, a). With evidence from 16 countries since the 1870s and across bond, equity, and housing markets, the return predictability of payout-price ratios, i.e., coupon price, dividend price, and rent price, remains weak in the sample, out of the sample, and economically. None of the 48 country-asset combinations shows consistent IS performance, OOS performance, and positive utility gain. When we limit our consideration to the bond and the equity markets, only two bond markets and four equity markets show IS predictability. The OOS performance is even worse, and the economic performance is terrible.

Using the weighted averages of excess returns from the bond, the equity, and the housing markets, I evaluate the predictability of the excess returns from the representative agents' portfolios in all 16 countries. The results are consistent with the findings from the individual markets, suggesting that the mean-variance investors cannot benefit from timing the representative agents' portfolios using the payout-price ratios.

In the end, I emphasize the findings of dividend predictability and rent predictability. Cochrane (2008, 2011) shows the relation between the return predictability and the dividend predictability and argues that the absence of the dividend predictability implies the return predictability. Consistent with the findings using data from the 1970s by Rangvid, Schmeling and Schrimpf (2014), I document that, in the past 150 years, 14 (5) countries show dividend (rent) predictability. Even the dividend growth in the US equity market is predictable from the 1870s to 2020. Taking together with Cochrane's VAR analysis, the IS and OOS tests, the findings from this paper collectively highlight the absence of the return predictability based on payout-price ratios across assets and countries.

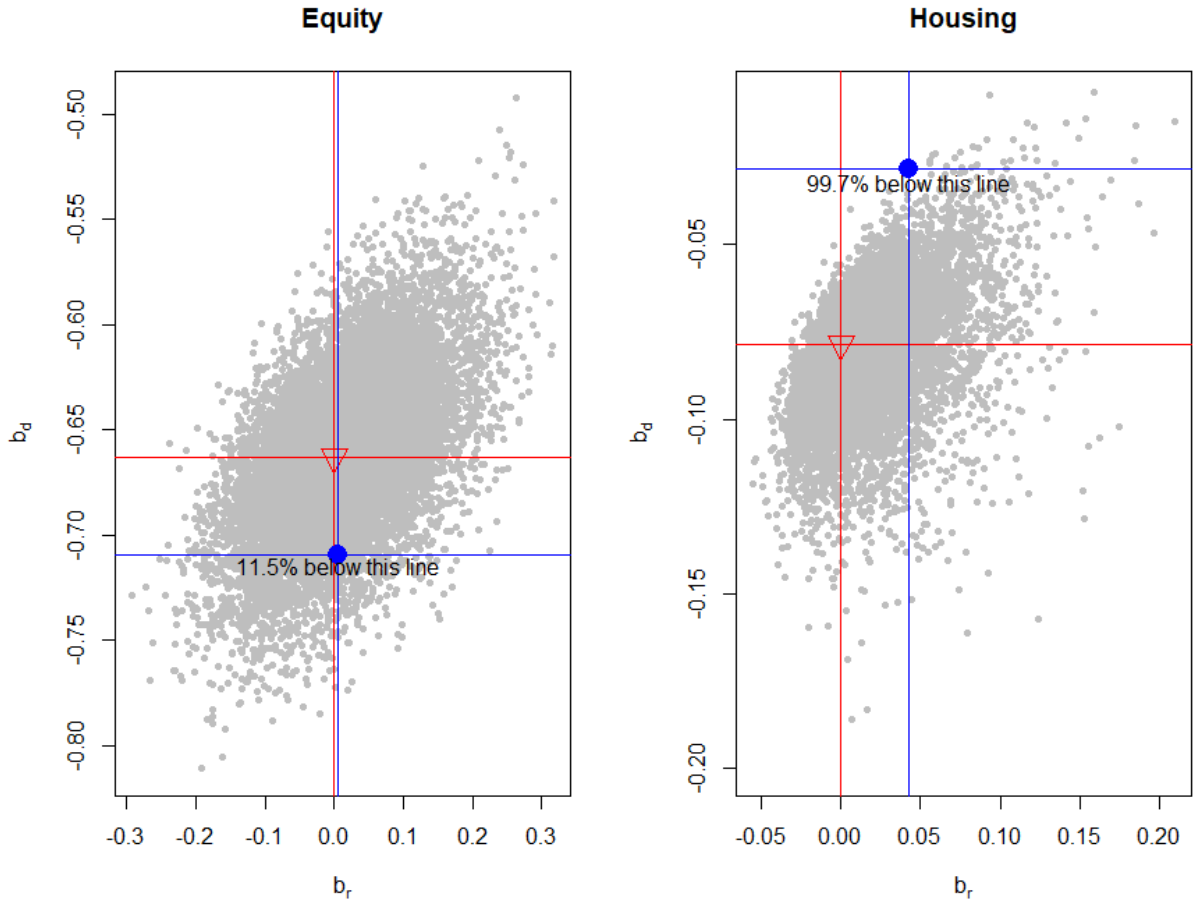

**Figure 1 VAR Simulation of Cochrane (2008)**

| | Estimates | | | | Null | | |
|---|---|---|---|---|---|---|---|
| | $\hat{\rho}$ | $\hat{\phi}$ | $\hat{b}_d$ | $\hat{b}_r$ | $\phi_0$ | $b_{d,0}$ | $b_{r,0}$ |
| Equity | 0.966 | 0.349 | -0.711 | 0.006 | 0.349 | -0.663 | 0.000 |
| Housing | 0.954 | 0.966 | -0.033 | 0.043 | 0.966 | -0.079 | 0.000 |

This figure presents the simulation results of excess return system following the VAR analysis by Cochrane (2008). Red lines and triangles mark the null hypothesis. Blue lines and solid dots mark the empirical estimates. The parameters are based on all the observations across the countries. The input of the covariance is based on the empirical estimates. The empirical estimates and the null hypotheses are stated in the above table. The simulation does not reject the hypotheses for absence of return predictability in both the equity and the housing markets. The null hypothesis of rent predictability in the housing markets is rejected. However, it is still possible that the coefficient of rent price is negative. The null hypothesis of the dividend predictability is not rejected, providing new evidence that the equity markets globally are not predictable with dividend price.

## Table 1 Summary of Results

This table reports the summary of in-sample and out-of-sample results across the countries and assets. The tests are all conducted for excess returns calculated as the difference between the asset returns and the treasury bills. The predictors are payout-price ratios, i.e., coupon-price ratios, dividend-price ratios, and the rent-price ratios. The tests cover long-term government bonds, stocks, housing assets, and representative portfolios formed with these assets. The risky portfolios include housing assets and stocks. The wealth portfolios include long-term government bonds, stocks, housing assets, and treasury bills. The representative agents' portfolios are value-weighted based on market capitalization. The predictors for representative agents' portfolios include all payout-price ratios. The "IS" columns report the in-sample significance of the predictors based on the Newey-West t stats. The "OOS" columns report the out-of-sample R-squared statistics. The "CER" columns report the Certainty Equivalent Return (CER) gains calculated using the null portfolios and the alternative portfolios, where the null portfolios are the prevailing average of excess returns, and the alternative portfolios are based on the predictions from the payout-price ratios. "Y" indicates the predictor successfully passed the test and marks either a significant in-sample coefficient, a significant positive out-of-sample R-squared value, or a positive CER gain. Predictions that pass both the in-sample test and the out-of-sample test are highlighted with gold background.

| Country | Bond | | | Equity | | | Housing | | | Risky | | | Wealth | | |
| --- | --- | --- | --- | --- | --- | --- | --- | --- | --- | --- | --- | --- | --- | --- | --- |
| | IS | OOS | CER | IS | OOS | CER | IS | OOS | CER | IS | OOS | CER | IS | OOS | CER |
| Australia | | | | Y | | Y | | | | | | | | | |
| Belgium | | | | Y | | | | | | | | | | | |
| Denmark | | | | | | | Y | | | Y | | | Y | | |
| Finland | | | | | | | Y | | Y | Y | | | Y | | |
| France | | | | Y | | | Y | Y | | Y | Y | | Y | Y | |
| Germany | | | | | | | Y | Y | | Y | | | Y | | |
| Italy | | | | | | | | | Y | Y | | Y | Y | | Y |
| Japan | Y | | Y | | | | Y | | Y | | | | | | |
| Netherlands | | | | | | Y | Y | | | Y | | | Y | | |
| Norway | | | | | | | | | | Y | | | Y | | |
| Portugal | Y | | | Y | | | Y | | | Y | | | Y | | |
| Spain | | | Y | | | Y | | | | Y | | | Y | | |
| Sweden | | | | | | | | | | Y | | | | | |
| Switzerland | | | | | | | | | Y | Y | | | Y | | |
| UK | | | | Y | Y | | Y | | | Y | | | Y | | |
| USA | | | | | | | | | | Y | | | Y | | |

**Table 2 In-Sample Predictive Regression**

This table reports the in-sample predictive OLS regressions of the excess returns from different asset classes using payout-price ratios in annual frequency. The excess returns are calculated as the difference between the asset returns and the treasury bill rates. The tests cover long-term government bonds, stocks, housing assets, and the two representative portfolio returns formed with these assets. The risky portfolios are the countries' representative agents' portfolios formed with housing assets and stocks. The wealth portfolios are the countries' representative agents' portfolios formed with bonds, stocks, housing assets, and treasury bills. These aggregated portfolios are value-weighted. The coefficients are the regression loadings on the predictor ratios. The t stats are the Newey-West t stats. Panel A through E report the results for bond excess returns, equity excess returns, housing excess returns, risky portfolio excess returns, and wealth portfolio excess returns, respectively.

| Panel A: Bond Excess Return on Coupon Price | | | | | |
|---|---|---|---|---|---|
| Country | Start | End | Coeff | NW t | $R^2$ |
| Australia | 1901 | 2020 | 0.024 | 1.626 | 0.022 |
| Belgium | 1871 | 2020 | 0.003 | 0.183 | 0.000 |
| Denmark | 1871 | 2020 | 0.007 | 0.474 | 0.003 |
| Finland | 1871 | 2020 | 0.016 | 1.369 | 0.016 |
| France | 1871 | 2020 | -0.006 | -0.687 | 0.002 |
| Germany | 1871 | 2020 | -0.001 | -0.207 | 0.000 |
| Italy | 1871 | 2020 | -0.005 | -0.255 | 0.001 |
| Japan | 1882 | 2020 | **-0.017** | -4.645 | 0.079 |
| Netherlands | 1871 | 2020 | -0.002 | -0.211 | 0.000 |
| Norway | 1871 | 2020 | 0.003 | 0.197 | 0.000 |
| Portugal | 1872 | 2020 | **0.026** | 1.652 | 0.019 |
| Spain | 1901 | 2020 | -0.015 | -1.009 | 0.013 |
| Sweden | 1872 | 2020 | 0.008 | 0.625 | 0.004 |
| Switzerland | 1901 | 2020 | -0.004 | -0.728 | 0.003 |
| UK | 1871 | 2020 | 0.016 | 1.183 | 0.009 |
| USA | 1872 | 2020 | 0.009 | 0.488 | 0.003 |

| Panel B: Equity Excess Return on Dividend Price | | | | | |
|---|---|---|---|---|---|
| Country | Start | End | Coeff | NW t | $R^2$ |
| Australia | 1871 | 2020 | **0.124** | 1.673 | 0.031 |
| Belgium | 1871 | 2020 | **0.049** | 2.021 | 0.013 |
| Denmark | 1874 | 2020 | -0.004 | -0.138 | 0.000 |
| Finland | 1913 | 2020 | 0.029 | 0.717 | 0.003 |
| France | 1871 | 2020 | **0.095** | 3.179 | 0.051 |
| Germany | 1871 | 2020 | -0.003 | -0.051 | 0.000 |
| Italy | 1871 | 2020 | 0.019 | 0.377 | 0.001 |
| Japan | 1887 | 2020 | 0.018 | 0.743 | 0.005 |
| Netherlands | 1901 | 2020 | 0.066 | 1.108 | 0.018 |
| Norway | 1882 | 2020 | 0.041 | 0.738 | 0.007 |
| Portugal | 1872 | 2020 | **0.029** | 7.066 | 0.147 |
| Spain | 1901 | 2020 | 0.001 | 0.659 | 0.000 |
| Sweden | 1872 | 2020 | 0.019 | 0.370 | 0.001 |
| Switzerland | 1901 | 2020 | -0.018 | -0.399 | 0.002 |
| UK | 1872 | 2020 | **0.216** | 3.878 | 0.129 |
| USA | 1873 | 2020 | 0.025 | 0.804 | 0.004 |

**Table 2 (Continued)**

| Panel C: Housing Excess Return on Rent Price | | | | | |
|---|---|---|---|---|---|
| Country | Start | End | Coeff | NW t | $R^2$ |
| Australia | 1902 | 2020 | 0.034 | 1.373 | 0.010 |
| Belgium | 1891 | 2020 | -0.033 | -1.270 | 0.013 |
| Denmark | 1877 | 2020 | **0.039** | 2.008 | 0.048 |
| Finland | 1921 | 2020 | **0.097** | 2.119 | 0.120 |
| France | 1872 | 2020 | **0.089** | 2.132 | 0.059 |
| Germany | 1872 | 2020 | **0.054** | 1.965 | 0.047 |
| Italy | 1929 | 2020 | 0.033 | 0.880 | 0.038 |
| Japan | 1932 | 2020 | **0.100** | 2.421 | 0.132 |
| Netherlands | 1872 | 2020 | **0.097** | 3.481 | 0.115 |
| Norway | 1872 | 2020 | 0.053 | 1.390 | 0.021 |
| Portugal | 1949 | 2020 | **0.086** | 2.533 | 0.103 |
| Spain | 1902 | 2020 | 0.048 | 1.601 | 0.028 |
| Sweden | 1884 | 2020 | 0.030 | 1.008 | 0.014 |
| Switzerland | 1903 | 2020 | 0.022 | 0.695 | 0.005 |
| UK | 1897 | 2020 | **0.071** | 2.501 | 0.037 |
| USA | 1892 | 2020 | 0.078 | 1.460 | 0.021 |

**Table 2 (Continued)**

| | | | Coupon Price | | Dividend Price | | Rent Price | | |
|---|---|---|---|---|---|---|---|---|---|
| Country | Start | End | Coeff | NW t | Coeff | NW t | Coeff | NW t | $R^2$ |
| Australia | 1902 | 2020 | -0.061 | -1.435 | 0.044 | 1.299 | -0.042 | -1.595 | 0.061 |
| Belgium | 1891 | 2020 | 0.004 | 0.178 | -0.002 | -0.044 | -0.015 | -0.955 | 0.005 |
| Denmark | 1880 | 2020 | -0.006 | -0.288 | 0.024 | 1.010 | **-0.026** | -2.510 | 0.064 |
| Finland | 1921 | 2019 | 0.018 | 0.483 | **0.118** | 2.613 | -0.011 | -0.551 | 0.110 |
| France | 1872 | 2020 | **-0.074** | -5.578 | **0.093** | 2.389 | -0.006 | -0.465 | 0.238 |
| Germany | 1872 | 2020 | -0.023 | -1.106 | **0.061** | 2.897 | **-0.028** | -3.254 | 0.091 |
| Italy | 1929 | 2020 | **-0.108** | -2.875 | **0.032** | 1.726 | -0.033 | -1.484 | 0.267 |
| Japan | 1932 | 2020 | 0.012 | 0.466 | 0.084 | 1.181 | 0.010 | 0.997 | 0.152 |
| Netherlands | 1901 | 2020 | -0.031 | -1.190 | **0.119** | 3.067 | -0.025 | -1.385 | 0.174 |
| Norway | 1882 | 2019 | -0.016 | -0.638 | **0.061** | 1.933 | -0.034 | -1.466 | 0.050 |
| Portugal | 1949 | 2019 | 0.001 | 0.605 | **0.134** | 4.065 | -0.020 | -0.810 | 0.181 |
| Spain | 1902 | 2017 | -0.020 | -0.884 | 0.032 | 1.214 | **-0.042** | -1.879 | 0.061 |
| Sweden | 1884 | 2019 | -0.026 | -0.974 | **0.060** | 1.729 | **-0.029** | -1.685 | 0.038 |
| Switzerland | 1903 | 2015 | -0.004 | -0.369 | 0.048 | 1.074 | **-0.035** | -3.584 | 0.079 |
| UK | 1897 | 2019 | **0.075** | 2.386 | **0.060** | 1.807 | -0.007 | -0.407 | 0.079 |
| USA | 1892 | 2020 | -0.004 | -0.206 | 0.074 | 1.406 | **-0.045** | -2.912 | 0.061 |

Panel D: Representative Agents' Risky Asset Portfolios on All Payout-Price Ratios

**Table 2 (Continued)**

| | | | Coupon Price | | Dividend Price | | Rent Price | | |
|---|---|---|---|---|---|---|---|---|---|
| Country | Start | End | Coeff | NW t | Coeff | NW t | Coeff | NW t | $R^2$ |
| Australia | 1902 | 2020 | -0.051 | -1.497 | 0.028 | 1.069 | -0.024 | -1.183 | 0.042 |
| Belgium | 1891 | 2020 | 0.003 | 0.163 | 0.004 | 0.128 | -0.012 | -0.811 | 0.005 |
| Denmark | 1880 | 2020 | -0.007 | -0.468 | 0.014 | 0.753 | **-0.020** | -2.399 | 0.047 |
| Finland | 1921 | 2019 | 0.017 | 0.516 | **0.108** | 2.688 | -0.006 | -0.319 | 0.116 |
| France | 1872 | 2020 | **-0.053** | -4.814 | **0.075** | 2.313 | -0.004 | -0.409 | 0.216 |
| Germany | 1872 | 2020 | -0.017 | -1.014 | **0.050** | 2.830 | **-0.021** | -3.084 | 0.087 |
| Italy | 1929 | 2020 | **-0.083** | -2.712 | 0.014 | 1.122 | -0.028 | -1.644 | 0.244 |
| Japan | 1932 | 2020 | 0.014 | 0.663 | 0.051 | 0.785 | 0.008 | 1.202 | 0.144 |
| Netherlands | 1901 | 2020 | -0.027 | -1.229 | **0.069** | 2.325 | -0.017 | -1.138 | 0.120 |
| Norway | 1882 | 2019 | -0.014 | -0.713 | **0.045** | 1.763 | **-0.031** | -1.779 | 0.054 |
| Portugal | 1949 | 2019 | 0.002 | 1.502 | **0.113** | 7.887 | -0.002 | -0.260 | 0.168 |
| Spain | 1902 | 2017 | -0.014 | -0.773 | 0.020 | 0.830 | **-0.035** | -1.895 | 0.059 |
| Sweden | 1884 | 2019 | -0.015 | -0.692 | 0.036 | 1.216 | -0.021 | -1.523 | 0.024 |
| Switzerland | 1903 | 2015 | -0.001 | -0.127 | 0.055 | 1.475 | **-0.033** | -3.935 | 0.090 |
| UK | 1897 | 2019 | **0.048** | 1.771 | 0.018 | 0.645 | 0.006 | 0.456 | 0.056 |
| USA | 1892 | 2020 | -0.008 | -0.547 | **0.085** | 1.944 | **-0.028** | -2.467 | 0.052 |

Panel E: Representative Agents' Wealth Portfolios on All Payout-Price Ratios

**Table 3 Out-of-Sample Tests**

This table reports the out-of-sample R-squared statistics in annual frequency with p values from Clark-West tests. The out-of-sample tests cover long-term government bonds, stocks, housing assets, and the two representative portfolio returns formed with these assets. For each asset class and each country, predictions start after 20 years after the data become available. The moving window is an expanding window, which uses all the historical data that are available. The risky portfolios are the countries' representative agents' portfolios formed with housing assets and stocks. The wealth portfolios are the countries' representative agents' portfolios formed with long-term government bonds, stocks, housing assets and treasury bills. These aggregated portfolios are value-weighted.

| Country | Bond $R^2_{OOS}$ | $P_{CW}$ | Equity $R^2_{OOS}$ | $P_{CW}$ | Housing $R^2_{OOS}$ | $P_{CW}$ | Risky $R^2_{OOS}$ | $P_{CW}$ | Wealth $R^2_{OOS}$ | $P_{CW}$ |
|---|---|---|---|---|---|---|---|---|---|---|
| Australia | -0.011 | 0.513 | 0.014 | 0.180 | -0.026 | 0.459 | -0.019 | 0.311 | -0.043 | 0.719 |
| Belgium | -0.030 | 0.547 | -0.010 | 0.317 | -0.100 | 0.628 | -0.318 | 0.655 | -0.249 | 0.548 |
| Denmark | -0.060 | 0.545 | -0.042 | 0.514 | -0.002 | 0.220 | -0.125 | 0.807 | -0.167 | 0.594 |
| Finland | -0.088 | 0.764 | -0.052 | 0.149 | -0.169 | 0.436 | -0.118 | 0.083 | -0.111 | 0.074 |
| France | -0.022 | 0.508 | 0.003 | 0.280 | **0.042** | 0.030 | **0.102** | 0.033 | **0.108** | 0.021 |
| Germany | -0.148 | 0.592 | -20.631 | 0.408 | **0.014** | 0.049 | -0.287 | 0.025 | -0.322 | 0.034 |
| Italy | -0.090 | 0.535 | -0.238 | 0.258 | 0.045 | 0.172 | -0.292 | 0.153 | -0.296 | 0.241 |
| Japan | 0.009 | 0.159 | -0.099 | 0.989 | -0.040 | 0.576 | -0.429 | 0.438 | -0.684 | 0.432 |
| Netherlands | -0.036 | 0.323 | -0.021 | 0.904 | 0.034 | 0.110 | -0.010 | 0.146 | -0.079 | 0.434 |
| Norway | -0.038 | 0.737 | -0.023 | 0.553 | -0.089 | 0.205 | -0.225 | 0.995 | -0.258 | 0.912 |
| Portugal | -0.049 | 0.828 | 0.030 | 0.313 | 0.052 | 0.144 | -4.703 | 0.601 | -4.891 | 0.670 |
| Spain | -0.065 | 0.371 | -0.547 | 0.990 | -0.017 | 0.454 | -0.115 | 0.150 | -0.093 | 0.129 |
| Sweden | -0.087 | 0.849 | -0.019 | 0.097 | -0.017 | 0.326 | -0.078 | 0.112 | -0.102 | 0.167 |
| Switzerland | -0.144 | 0.233 | -0.120 | 0.228 | -0.051 | 0.339 | -0.074 | 0.133 | -0.066 | 0.145 |
| UK | -0.050 | 0.204 | **0.099** | 0.020 | -0.065 | 0.019 | -0.151 | 0.543 | -0.186 | 0.672 |
| USA | -0.029 | 0.869 | -0.022 | 0.711 | -0.224 | 0.928 | -0.195 | 0.907 | -0.267 | 0.605 |

**Table 4 Out-of-Sample Economic Performance**

This table reports the Sharpe ratio, Certainty Equivalent Return (CER) gains, and relative turnover based on the null portfolios formed with the prevailing historical mean of excess returns and the alternative portfolios formed with the predictions made by payout-price ratios. Both the null portfolios and the alternative portfolios are constructed from the mean-variance investor's perspective. The risk aversion coefficient is assumed to be 5. Sharpe Ratios and relative turnover are in decimal, while CER gains are in percentage. Bold font indicates positive CER gain. The significance of the CERs is decided following DeMiguel, Garlappi, and Uppal (2009) and reported through z stats.

| Panel A: Bond Markets | | | | | |
|---|---|---|---|---|---|
| | Null SR | Alt SR | CER Gain | CER Z | Turnover |
| Australia | -0.01 | 0.05 | -0.47 | -1.47 | 2.03 |
| Belgium | -0.01 | 0.00 | -0.69 | -2.74 | 2.78 |
| Denmark | -0.02 | -0.06 | -0.69 | -4.13 | 5.91 |
| Finland | 0.12 | 0.20 | -0.90 | -1.01 | 6.71 |
| France | 0.00 | -0.02 | -0.25 | -9.64 | 2.16 |
| Germany | 0.27 | 0.22 | -0.40 | -9.28 | 1.83 |
| Italy | -0.09 | -0.03 | -0.50 | -1.06 | 5.28 |
| Japan | 0.00 | 0.19 | **0.66** | 7.32 | Inf |
| Netherlands | -0.05 | -0.14 | -0.58 | -12.35 | 2.75 |
| Norway | -0.05 | -0.13 | -0.44 | -5.95 | 2.68 |
| Portugal | 0.06 | 0.03 | -0.94 | -2.29 | 1.70 |
| Spain | -0.04 | 0.00 | **0.14** | 2.60 | 2.17 |
| Sweden | 0.01 | 0.08 | -0.30 | -1.30 | 4.82 |
| Switzerland | 0.41 | 0.32 | -0.46 | -6.99 | 2.69 |
| UK | 0.02 | -0.02 | -0.22 | -10.51 | 3.53 |
| USA | 0.01 | 0.02 | 0.01 | 0.29 | 54.04 |

**Table 4 (Continued)**

| | Null SR | Alt SR | CER Gain | CER Z | Turnover |
|---|---|---|---|---|---|
| Panel B: Equity Markets | | | | | |
| Australia | 0.40 | 0.42 | **0.27** | 6.27 | 1.53 |
| Belgium | 0.15 | 0.17 | -0.10 | -0.67 | 4.53 |
| Denmark | 0.21 | 0.16 | -0.78 | -4.14 | 1.95 |
| Finland | 0.30 | 0.26 | -0.71 | -3.94 | 3.23 |
| France | 0.15 | 0.18 | 0.02 | 0.08 | 4.77 |
| Germany | 0.09 | 0.09 | -6.05E+18 | 0.00 | 1.34 |
| Italy | 0.06 | 0.01 | -1.64 | -3.11 | 4.20 |
| Japan | 0.13 | 0.05 | -0.47 | -3.99 | 3.30 |
| Netherlands | 0.33 | 0.38 | **1.52** | 1.90 | 3.69 |
| Norway | 0.01 | 0.01 | -0.48 | -3.14 | 4.57 |
| Portugal | 0.20 | 0.14 | -1.11 | -3.81 | 2.32 |
| Spain | 0.16 | 0.22 | **0.34** | 2.99 | 2.30 |
| Sweden | 0.18 | 0.15 | -0.19 | -28.27 | 1.57 |
| Switzerland | 0.23 | 0.13 | -1.36 | -2.42 | 3.31 |
| UK | 0.26 | 0.31 | -0.29 | -0.44 | 3.31 |
| USA | 0.25 | 0.18 | -0.31 | -1.59 | 2.39 |

**Table 4 (Continued)**

| | Null SR | Alt SR | CER Gain | CER Z | Turnover |
|---|---|---|---|---|---|
| Panel C: Housing Markets | | | | | |
| Australia | 0.38 | 0.32 | -2.74 | -4.74 | 2.85 |
| Belgium | 0.81 | 0.79 | -0.48 | -1.64 | 4.09 |
| Denmark | 0.73 | 0.70 | -0.47 | -1.54 | 1.67 |
| Finland | 0.73 | 0.77 | **0.61** | 1.82 | 1.34 |
| France | 0.89 | 0.85 | -0.51 | -5.92 | 3.90 |
| Germany | 0.43 | 0.48 | 0.39 | 0.90 | 2.40 |
| Italy | 0.29 | 0.33 | **0.53** | 14.74 | 0.86 |
| Japan | 0.46 | 0.47 | **0.13** | 11.60 | Inf |
| Netherlands | 0.77 | 0.82 | 0.21 | 0.38 | 2.40 |
| Norway | 0.72 | 0.64 | -1.03 | -1.67 | 3.24 |
| Portugal | 0.78 | 0.77 | -0.22 | -4.57 | 4.26 |
| Spain | 0.40 | 0.36 | -0.20 | -0.65 | 2.28 |
| Sweden | 0.82 | 0.73 | -0.87 | -2.83 | 1.97 |
| Switzerland | 1.15 | 1.14 | **0.01** | 3.54 | 0.00 |
| UK | 0.55 | 0.55 | -0.12 | -0.21 | 3.31 |
| USA | 0.76 | 0.62 | -0.82 | -6.92 | 1.97 |

| Panel D: Representative Agents' Risky Asset Portfolios | | | | |
|---|---|---|---|---|
| | Null SR | Alt SR | CER Gain | CER Z | Turnover |
| Australia | 0.43 | 0.37 | -2.45 | -4.67 | 6.29 |
| Belgium | 0.64 | 0.59 | -0.83 | -1.43 | 5.07 |
| Denmark | 0.70 | 0.66 | -0.45 | -3.02 | 2.39 |
| Finland | 0.50 | 0.64 | 1.26 | 0.88 | 1.32 |
| France | 0.92 | 0.75 | -1.87 | -3.47 | 11.39 |
| Germany | 0.44 | 0.56 | 0.93 | 1.02 | 7.61 |
| Italy | 0.30 | 0.34 | **0.56** | 2.20 | 3.97 |
| Japan | 0.42 | 0.40 | -0.15 | -1.08 | 4.64 |
| Netherlands | 0.81 | 0.77 | -0.72 | -2.14 | 7.11 |
| Norway | 0.76 | 0.62 | -1.42 | -4.48 | 5.27 |
| Portugal | 0.65 | 0.56 | -1.60 | -5.98 | 2.89 |
| Spain | 0.46 | 0.46 | 0.08 | 0.13 | 5.11 |
| Sweden | 0.73 | 0.70 | -0.39 | -1.13 | 3.50 |
| Switzerland | 0.90 | 0.88 | -0.06 | -0.36 | 6.25 |
| UK | 0.57 | 0.48 | -0.95 | -1.00 | 3.88 |
| USA | 0.62 | 0.52 | -1.07 | -4.63 | 7.24 |

**Table 4 (Continued)**

| | Null SR | Alt SR | CER Gain | CER Z | Turnover |
|---|---|---|---|---|---|
| Panel E: Representative Agents' Wealth Portfolios | | | | | |
| Australia | 0.45 | 0.43 | -0.28 | -3.45 | 6.90 |
| Belgium | 0.64 | 0.54 | -1.02 | -2.16 | 4.77 |
| Denmark | 0.71 | 0.67 | -0.47 | -1.76 | 5.05 |
| Finland | 0.50 | 0.63 | 1.27 | 1.06 | 1.32 |
| France | 0.86 | 0.70 | -1.53 | -5.86 | 20.24 |
| Germany | 0.45 | 0.60 | 0.69 | 1.04 | 8.50 |
| Italy | 0.36 | 0.38 | **0.26** | 1.80 | 4.12 |
| Japan | 0.47 | 0.45 | -0.21 | -5.51 | 6.22 |
| Netherlands | 0.84 | 0.76 | -0.83 | -9.20 | 9.16 |
| Norway | 0.76 | 0.66 | -1.01 | -4.36 | 6.65 |
| Portugal | 0.63 | 0.51 | -1.85 | -7.76 | 2.73 |
| Spain | 0.50 | 0.45 | -0.49 | -1.04 | 5.50 |
| Sweden | 0.77 | 0.75 | -0.37 | -1.36 | 4.30 |
| Switzerland | 0.92 | 0.87 | -0.38 | -2.40 | 5.28 |
| UK | 0.53 | 0.43 | -0.89 | -1.50 | 3.09 |
| USA | 0.61 | 0.53 | -0.66 | -8.14 | 5.73 |

**Table 5 Payout Predictability: Dividend Growth and Rent Growth**

This table reports the OLS regressions of payout growth on payout-price ratios. Panel A reports for the equity markets using dividend growth and dividend price. Panel B reports for the housing markets using rent growth and rent price.

| Country | Start | End | Coeff | NW t | $R^2$ |
|---|---|---|---|---|---|
| Panel A: Dividend Growth on Dividend Price | | | | | |
| Australia | 1871 | 2020 | -0.088 | -1.317 | 0.017 |
| Belgium | 1871 | 2020 | **-0.125** | -2.097 | 0.045 |
| Denmark | 1874 | 2020 | **-0.119** | -2.847 | 0.060 |
| Finland | 1914 | 2020 | **-0.256** | -4.040 | 0.130 |
| France | 1871 | 2020 | -0.033 | -0.582 | 0.008 |
| Germany | 1871 | 2020 | **-0.871** | -40.747 | 0.904 |
| Italy | 1871 | 2020 | **-0.175** | -3.410 | 0.091 |
| Japan | 1887 | 2020 | **-0.028** | -1.779 | 0.024 |
| Netherlands | 1901 | 2020 | **-0.366** | -3.740 | 0.192 |
| Norway | 1882 | 2020 | **-0.275** | -2.044 | 0.117 |
| Portugal | 1872 | 2020 | **-0.695** | -4.702 | 0.635 |
| Spain | 1901 | 2020 | **-0.864** | -28.774 | 0.865 |
| Sweden | 1872 | 2020 | **-0.296** | -5.682 | 0.201 |
| Switzerland | 1901 | 2020 | **-0.194** | -4.621 | 0.128 |
| UK | 1872 | 2020 | **-0.164** | -1.705 | 0.067 |
| USA | 1873 | 2020 | **-0.097** | -3.133 | 0.127 |

**Table 5 (Continued)**

| Country | Start | End | Coeff | NW t | $R^2$ |
|---------|-------|-----|-------|------|-------|
| Panel B: Rent Growth on Rent Price | | | | | |
| Australia | 1902 | 2020 | 0.028 | 1.107 | 0.007 |
| Belgium | 1891 | 2020 | **-0.097** | -1.827 | 0.191 |
| Denmark | 1877 | 2020 | -0.028 | -1.613 | 0.089 |
| Finland | 1921 | 2020 | -0.010 | -0.174 | 0.001 |
| France | 1872 | 2020 | 0.050 | 1.272 | 0.020 |
| Germany | 1872 | 2020 | **-0.064** | -1.831 | 0.083 |
| Italy | 1929 | 2020 | -0.033 | -0.795 | 0.044 |
| Japan | 1932 | 2020 | -0.020 | -0.489 | 0.007 |
| Netherlands | 1872 | 2020 | 0.004 | 0.273 | 0.002 |
| Norway | 1872 | 2020 | -0.011 | -0.463 | 0.002 |
| Portugal | 1949 | 2020 | **0.080** | 2.630 | 0.112 |
| Spain | 1902 | 2020 | -0.008 | -0.388 | 0.002 |
| Sweden | 1884 | 2020 | 0.001 | 0.051 | 0.000 |
| Switzerland | 1903 | 2020 | **-0.070** | -3.047 | 0.165 |
| UK | 1897 | 2020 | -0.020 | -0.673 | 0.007 |
| USA | 1892 | 2020 | **-0.129** | -4.208 | 0.233 |

# Appendix

**Table A1 Summary Statistics**

This table reports the summary statistics of the variables in this paper. Panel A (B/C/D) reports the summary statistics of the bond markets (equity markets/housing markets/representative agents' portfolios) with coupon-price ratio (dividend-price ratio/rent-price ratio/weighted average of excess returns). The excess returns and payout-price ratios are all in log scales.

| Panel A: Bond Markets | | | | | | | | |
|---|---|---|---|---|---|---|---|---|
| Country | Variable | Mean | SD | Min | Q1 | Median | Q3 | Max |
| Australia | Bond Excess Return | 0.01 | 0.07 | -0.18 | -0.03 | 0.00 | 0.04 | 0.23 |
| Australia | CP | -2.97 | 0.41 | -3.77 | -3.26 | -3.00 | -2.80 | -1.97 |
| Belgium | Bond Excess Return | 0.01 | 0.07 | -0.20 | -0.03 | 0.01 | 0.05 | 0.25 |
| Belgium | CP | -3.13 | 0.45 | -4.95 | -3.39 | -3.15 | -2.85 | -2.05 |
| Denmark | Bond Excess Return | 0.00 | 0.07 | -0.22 | -0.04 | 0.00 | 0.04 | 0.37 |
| Denmark | CP | -3.07 | 0.59 | -5.79 | -3.30 | -3.14 | -2.88 | -1.71 |
| Finland | Bond Excess Return | 0.01 | 0.09 | -0.28 | -0.02 | 0.01 | 0.05 | 0.35 |
| Finland | CP | -2.92 | 0.71 | -7.29 | -3.15 | -2.99 | -2.54 | -1.57 |
| France | Bond Excess Return | 0.01 | 0.09 | -0.25 | -0.04 | 0.01 | 0.07 | 0.28 |
| France | CP | -3.14 | 0.60 | -6.72 | -3.46 | -3.13 | -2.80 | -1.85 |
| Germany | Bond Excess Return | 0.02 | 0.05 | -0.19 | -0.01 | 0.01 | 0.04 | 0.32 |
| Germany | CP | -3.13 | 0.61 | -7.04 | -3.31 | -3.10 | -2.79 | -2.34 |
| Italy | Bond Excess Return | 0.01 | 0.09 | -0.34 | -0.02 | 0.01 | 0.04 | 0.32 |
| Italy | CP | -2.84 | 0.43 | -4.22 | -3.06 | -2.91 | -2.69 | -1.67 |
| Japan | Bond Excess Return | 0.00 | 0.05 | -0.15 | -0.03 | -0.01 | 0.03 | 0.19 |
| Japan | CP | -3.22 | 0.86 | -7.57 | -3.20 | -2.87 | -2.81 | -2.27 |
| Netherlands | Bond Excess Return | 0.01 | 0.06 | -0.21 | -0.03 | 0.01 | 0.05 | 0.20 |
| Netherlands | CP | -3.25 | 0.50 | -5.93 | -3.46 | -3.30 | -3.01 | -2.21 |
| Norway | Bond Excess Return | 0.00 | 0.05 | -0.18 | -0.02 | 0.00 | 0.02 | 0.18 |
| Norway | CP | -3.14 | 0.45 | -4.66 | -3.38 | -3.13 | -2.94 | -2.03 |
| Portugal | Bond Excess Return | 0.02 | 0.11 | -0.38 | -0.02 | 0.02 | 0.07 | 0.58 |
| Portugal | CP | -2.87 | 0.55 | -4.94 | -3.24 | -2.99 | -2.39 | -1.80 |
| Spain | Bond Excess Return | 0.01 | 0.07 | -0.25 | -0.01 | 0.01 | 0.03 | 0.25 |
| Spain | CP | -3.05 | 0.50 | -5.13 | -3.22 | -3.13 | -3.02 | -1.87 |
| Sweden | Bond Excess Return | 0.01 | 0.08 | -0.42 | -0.03 | 0.00 | 0.03 | 0.21 |
| Sweden | CP | -3.18 | 0.64 | -7.73 | -3.37 | -3.20 | -2.97 | -2.12 |
| Switzerland | Bond Excess Return | 0.02 | 0.05 | -0.13 | -0.01 | 0.01 | 0.04 | 0.16 |
| Switzerland | CP | -3.40 | 0.61 | -8.02 | -3.49 | -3.31 | -3.11 | -2.61 |
| UK | Bond Excess Return | 0.01 | 0.09 | -0.26 | -0.04 | 0.00 | 0.04 | 0.31 |
| UK | CP | -3.20 | 0.52 | -4.73 | -3.56 | -3.30 | -2.92 | -1.71 |
| USA | Bond Excess Return | 0.00 | 0.07 | -0.18 | -0.03 | 0.00 | 0.03 | 0.22 |
| USA | CP | -3.24 | 0.42 | -4.05 | -3.46 | -3.30 | -3.03 | -1.97 |

**Table A1 (Continued)**

| Country | Variable | Mean | SD | Min | Q1 | Median | Q3 | Max |
|---------|----------|------|-----|-----|-----|--------|-----|-----|
| | | | | Panel B: Equity Markets | | | | |
| Australia | Equity Excess Return | 0.05 | 0.14 | -0.58 | 0.00 | 0.07 | 0.13 | 0.36 |
| Australia | DP | -3.03 | 0.20 | -3.56 | -3.19 | -3.03 | -2.88 | -2.59 |
| Belgium | Equity Excess Return | 0.03 | 0.20 | -0.86 | -0.08 | 0.01 | 0.13 | 0.79 |
| Belgium | DP | -3.37 | 0.47 | -5.81 | -3.52 | -3.26 | -3.11 | -2.44 |
| Denmark | Equity Excess Return | 0.04 | 0.16 | -0.70 | -0.05 | 0.02 | 0.10 | 0.46 |
| Denmark | DP | -3.18 | 0.55 | -4.59 | -3.35 | -2.98 | -2.82 | -2.17 |
| Finland | Equity Excess Return | 0.07 | 0.26 | -0.76 | -0.08 | 0.07 | 0.23 | 0.95 |
| Finland | DP | -3.14 | 0.48 | -4.78 | -3.31 | -3.03 | -2.85 | -2.34 |
| France | Equity Excess Return | 0.02 | 0.19 | -0.56 | -0.08 | 0.02 | 0.14 | 0.73 |
| France | DP | -3.40 | 0.44 | -5.35 | -3.54 | -3.29 | -3.17 | -2.57 |
| Germany | Equity Excess Return | 0.20 | 1.80 | -2.19 | -0.06 | 0.05 | 0.15 | 21.64 |
| Germany | DP | -3.87 | 3.21 | -25.72 | -3.69 | -3.34 | -3.00 | -2.06 |
| Italy | Equity Excess Return | 0.03 | 0.25 | -0.68 | -0.09 | 0.02 | 0.16 | 0.91 |
| Italy | DP | -3.35 | 0.48 | -5.57 | -3.55 | -3.19 | -3.05 | -2.60 |
| Japan | Equity Excess Return | 0.02 | 0.20 | -0.52 | -0.12 | 0.03 | 0.15 | 0.61 |
| Japan | DP | -3.51 | 0.82 | -5.41 | -4.21 | -3.21 | -2.85 | -2.33 |
| Netherlands | Equity Excess Return | 0.06 | 0.18 | -0.72 | -0.04 | 0.07 | 0.17 | 0.54 |
| Netherlands | DP | -3.14 | 0.37 | -4.01 | -3.44 | -3.16 | -2.86 | -2.30 |
| Norway | Equity Excess Return | 0.02 | 0.18 | -0.82 | -0.06 | 0.02 | 0.13 | 0.52 |
| Norway | DP | -3.27 | 0.39 | -4.57 | -3.46 | -3.24 | -3.04 | -2.12 |
| Portugal | Equity Excess Return | 0.02 | 0.22 | -0.73 | -0.06 | 0.03 | 0.10 | 0.67 |
| Portugal | DP | -3.94 | 2.91 | -25.72 | -3.81 | -3.34 | -2.96 | -2.63 |
| Spain | Equity Excess Return | 0.04 | 0.18 | -0.50 | -0.08 | 0.04 | 0.15 | 0.65 |
| Spain | DP | -3.80 | 3.55 | -25.72 | -3.52 | -3.24 | -2.99 | -2.19 |
| Sweden | Equity Excess Return | 0.04 | 0.18 | -0.53 | -0.05 | 0.06 | 0.14 | 0.49 |
| Sweden | DP | -3.26 | 0.32 | -4.23 | -3.41 | -3.23 | -3.05 | -2.61 |
| Switzerland | Equity Excess Return | 0.04 | 0.18 | -0.44 | -0.06 | 0.05 | 0.15 | 0.45 |
| Switzerland | DP | -3.57 | 0.50 | -4.88 | -3.93 | -3.52 | -3.20 | -2.23 |
| UK | Equity Excess Return | 0.04 | 0.16 | -0.79 | -0.02 | 0.04 | 0.13 | 0.81 |
| UK | DP | -3.21 | 0.27 | -3.91 | -3.40 | -3.20 | -3.04 | -2.15 |
| USA | Equity Excess Return | 0.05 | 0.17 | -0.54 | -0.06 | 0.07 | 0.16 | 0.40 |
| USA | DP | -3.23 | 0.45 | -4.45 | -3.45 | -3.15 | -2.93 | -2.29 |

**Table A1 (Continued)**

| Country | Variable | Mean | SD | Min | Q1 | Median | Q3 | Max |
|---------|----------|------|-----|-----|-----|--------|-----|-----|
| | | Panel C: Housing Markets | | | | | | |
| Australia | Housing Excess Return | 0.05 | 0.10 | -0.20 | 0.00 | 0.04 | 0.08 | 0.84 |
| Australia | RP | -3.33 | 0.28 | -4.11 | -3.42 | -3.26 | -3.13 | -2.85 |
| Belgium | Housing Excess Return | 0.07 | 0.09 | -0.15 | 0.03 | 0.06 | 0.09 | 0.38 |
| Belgium | RP | -2.89 | 0.30 | -3.49 | -3.13 | -2.82 | -2.68 | -2.36 |
| Denmark | Housing Excess Return | 0.05 | 0.07 | -0.16 | 0.02 | 0.06 | 0.09 | 0.24 |
| Denmark | RP | -2.77 | 0.41 | -3.71 | -3.12 | -2.60 | -2.43 | -2.14 |
| Finland | Housing Excess Return | 0.07 | 0.13 | -0.27 | 0.04 | 0.08 | 0.13 | 0.53 |
| Finland | RP | -2.81 | 0.45 | -3.87 | -2.94 | -2.76 | -2.44 | -2.18 |
| France | Housing Excess Return | 0.07 | 0.08 | -0.10 | 0.02 | 0.06 | 0.11 | 0.42 |
| France | RP | -3.08 | 0.23 | -3.67 | -3.18 | -3.06 | -2.92 | -2.65 |
| Germany | Housing Excess Return | 0.04 | 0.10 | -0.34 | 0.00 | 0.04 | 0.09 | 0.44 |
| Germany | RP | -2.93 | 0.38 | -3.50 | -3.26 | -2.98 | -2.71 | -2.03 |
| Italy | Housing Excess Return | 0.03 | 0.11 | -0.16 | -0.02 | 0.02 | 0.07 | 0.60 |
| Italy | RP | -3.57 | 0.66 | -5.31 | -3.58 | -3.39 | -3.18 | -2.82 |
| Japan | Housing Excess Return | 0.05 | 0.07 | -0.13 | 0.00 | 0.04 | 0.08 | 0.28 |
| Japan | RP | -3.15 | 0.26 | -3.67 | -3.38 | -3.15 | -2.95 | -2.60 |
| Netherlands | Housing Excess Return | 0.06 | 0.09 | -0.25 | 0.01 | 0.06 | 0.11 | 0.29 |
| Netherlands | RP | -2.89 | 0.32 | -3.61 | -3.10 | -2.91 | -2.67 | -2.04 |
| Norway | Housing Excess Return | 0.05 | 0.08 | -0.23 | 0.01 | 0.06 | 0.11 | 0.42 |
| Norway | RP | -2.76 | 0.23 | -3.42 | -2.84 | -2.70 | -2.61 | -2.43 |
| Portugal | Housing Excess Return | 0.07 | 0.08 | -0.19 | 0.02 | 0.08 | 0.12 | 0.28 |
| Portugal | RP | -3.28 | 0.32 | -3.77 | -3.50 | -3.33 | -3.07 | -2.38 |
| Spain | Housing Excess Return | 0.05 | 0.12 | -0.36 | -0.03 | 0.04 | 0.12 | 0.38 |
| Spain | RP | -3.34 | 0.40 | -4.21 | -3.61 | -3.30 | -3.07 | -2.40 |
| Sweden | Housing Excess Return | 0.06 | 0.07 | -0.30 | 0.01 | 0.07 | 0.10 | 0.27 |
| Sweden | RP | -2.73 | 0.29 | -3.59 | -2.87 | -2.68 | -2.53 | -2.16 |
| Switzerland | Housing Excess Return | 0.05 | 0.06 | -0.09 | 0.01 | 0.06 | 0.09 | 0.20 |
| Switzerland | RP | -3.12 | 0.19 | -3.51 | -3.26 | -3.13 | -2.98 | -2.73 |
| UK | Housing Excess Return | 0.04 | 0.09 | -0.18 | -0.01 | 0.04 | 0.09 | 0.29 |
| UK | RP | -3.32 | 0.24 | -3.81 | -3.46 | -3.30 | -3.14 | -2.78 |
| USA | Housing Excess Return | 0.05 | 0.08 | -0.31 | 0.02 | 0.04 | 0.08 | 0.32 |
| USA | RP | -2.98 | 0.15 | -3.30 | -3.06 | -2.99 | -2.92 | -2.57 |

**Table A1 (Continued)**

| Country | Variable | Mean | SD | Min | Q1 | Median | Q3 | Max |
|---|---|---|---|---|---|---|---|---|
| | Panel D: Representative Agents' Portfolios | | | | | | | |
| Australia | Risky Excess Return | 0.05 | 0.09 | -0.23 | 0.02 | 0.05 | 0.08 | 0.76 |
| | Wealth Excess Return | 0.04 | 0.07 | -0.15 | 0.02 | 0.04 | 0.07 | 0.60 |
| Belgium | Risky Excess Return | 0.07 | 0.11 | -0.19 | 0.01 | 0.05 | 0.11 | 0.51 |
| | Wealth Excess Return | 0.05 | 0.08 | -0.10 | 0.01 | 0.04 | 0.08 | 0.41 |
| Denmark | Risky Excess Return | 0.05 | 0.08 | -0.15 | 0.02 | 0.05 | 0.09 | 0.25 |
| | Wealth Excess Return | 0.04 | 0.06 | -0.12 | 0.01 | 0.04 | 0.08 | 0.23 |
| Finland | Risky Excess Return | 0.08 | 0.16 | -0.25 | 0.02 | 0.08 | 0.14 | 0.80 |
| | Wealth Excess Return | 0.07 | 0.14 | -0.22 | 0.02 | 0.07 | 0.12 | 0.73 |
| France | Risky Excess Return | 0.07 | 0.08 | -0.07 | 0.01 | 0.05 | 0.11 | 0.35 |
| | Wealth Excess Return | 0.05 | 0.06 | -0.08 | 0.01 | 0.04 | 0.09 | 0.24 |
| Germany | Risky Excess Return | 0.04 | 0.09 | -0.35 | 0.00 | 0.04 | 0.08 | 0.39 |
| | Wealth Excess Return | 0.03 | 0.07 | -0.26 | 0.00 | 0.03 | 0.07 | 0.35 |
| Italy | Risky Excess Return | 0.04 | 0.11 | -0.18 | -0.02 | 0.03 | 0.07 | 0.57 |
| | Wealth Excess Return | 0.04 | 0.09 | -0.09 | 0.00 | 0.02 | 0.06 | 0.47 |
| Japan | Risky Excess Return | 0.05 | 0.08 | -0.12 | 0.01 | 0.04 | 0.08 | 0.27 |
| | Wealth Excess Return | 0.04 | 0.06 | -0.11 | 0.01 | 0.04 | 0.07 | 0.22 |
| Netherlands | Risky Excess Return | 0.07 | 0.09 | -0.15 | 0.02 | 0.06 | 0.12 | 0.33 |
| | Wealth Excess Return | 0.05 | 0.07 | -0.12 | 0.01 | 0.05 | 0.09 | 0.21 |
| Norway | Risky Excess Return | 0.05 | 0.07 | -0.19 | 0.02 | 0.06 | 0.10 | 0.28 |
| | Wealth Excess Return | 0.04 | 0.06 | -0.16 | 0.01 | 0.05 | 0.08 | 0.23 |
| Portugal | Risky Excess Return | 0.07 | 0.09 | -0.24 | 0.03 | 0.08 | 0.13 | 0.31 |
| | Wealth Excess Return | 0.06 | 0.09 | -0.23 | 0.03 | 0.07 | 0.12 | 0.30 |
| Spain | Risky Excess Return | 0.05 | 0.10 | -0.33 | -0.02 | 0.05 | 0.10 | 0.34 |
| | Wealth Excess Return | 0.04 | 0.08 | -0.21 | -0.01 | 0.04 | 0.09 | 0.25 |
| Sweden | Risky Excess Return | 0.06 | 0.08 | -0.14 | 0.00 | 0.06 | 0.11 | 0.37 |
| | Wealth Excess Return | 0.05 | 0.07 | -0.13 | 0.00 | 0.05 | 0.09 | 0.30 |
| Switzerland | Risky Excess Return | 0.05 | 0.07 | -0.11 | 0.00 | 0.06 | 0.11 | 0.24 |
| | Wealth Excess Return | 0.05 | 0.06 | -0.09 | 0.00 | 0.05 | 0.09 | 0.21 |
| UK | Risky Excess Return | 0.05 | 0.09 | -0.19 | -0.01 | 0.06 | 0.11 | 0.27 |
| | Wealth Excess Return | 0.03 | 0.07 | -0.13 | -0.01 | 0.04 | 0.08 | 0.23 |
| USA | Risky Excess Return | 0.05 | 0.09 | -0.34 | 0.01 | 0.06 | 0.10 | 0.31 |
| | Wealth Excess Return | 0.04 | 0.08 | -0.33 | 0.01 | 0.05 | 0.08 | 0.30 |